\pdfoutput=1

\documentclass[10pt,prd,aps,amssymb,amsmath,tightenlines,showpacs]{revtex4}
\usepackage{graphicx}

\newcommand{\cC}{\ensuremath{\mathcal{C}}}
\newcommand{\cB}{\ensuremath{\mathcal{B}}}
\newcommand{\cM}{\ensuremath{\mathcal{M}}}
\newcommand{\cN}{\ensuremath{\mathcal{N}}}

\newcommand{\cP}{\ensuremath{\mathcal{P}}}
\newcommand{\cT}{\ensuremath{\mathcal{T}}}
\newcommand{\cPT}{\ensuremath{\mathcal{PT}}}

\newcommand{\half}{\mbox{$\textstyle{\frac{1}{2}}$}}

\begin{document}

\title{Nonuniqueness of the $\cC$ operator in $\cP\cT$-symmetric quantum
mechanics}

\author{Carl M. Bender$^a$}\email{cmb@wustl.edu}
\author{Mariagiovanna Gianfreda$^b$}\email{Maria.Gianfreda@le.infn.it}
\affiliation{$^a$Department of Physics, Washington University, St. Louis, MO
63130, USA\\ $^b$Dipartimento di Matematica e Fisica Ennio De Giorgi,
Universit\`a del Salento and I.N.F.N. Sezione di Lecce, Via Arnesano, I-73100
Lecce, Italy}

\date{\today}

\begin{abstract}
The $\cC$ operator in $\cPT$-symmetric quantum mechanics satisfies a system of
three simultaneous algebraic operator equations, $\cC^2=1$, $[\cC,\cPT]=0$, and
$[\cC,H]=0$. These equations are difficult to solve exactly, so perturbative
methods have been used in the past to calculate $\cC$. The usual approach has
been to express the Hamiltonian as $H=H_0+\epsilon H_1$, and to seek a solution
for $\cC$ in the form $\cC=e^Q\cP$, where $Q=Q(q,p)$ is odd in the momentum $p$,
even in the coordinate $q$, and has a perturbation expansion of the form $Q=
\epsilon Q_1+\epsilon^3 Q_3+\epsilon^5 Q_5+\ldots$. [In previous work it has
always been assumed that the coefficients of even powers of $\epsilon$ in this
expansion would be absent because their presence would violate the condition
that $Q(p,q)$ is odd in $p$.] In an earlier paper it was argued that the $\cC$
operator is not unique because the perturbation coefficient $Q_1$ is nonunique.
Here, the nonuniqueness of $\cC$ is demonstrated at a more fundamental level: It
is shown that the perturbation expansion for $Q$ actually has the more general
form $Q=Q_0+\epsilon Q_1+\epsilon^2 Q_2+\ldots$ in which {\it all} powers and
not just odd powers of $\epsilon$ appear. For the case in which $H_0$ is the
harmonic-oscillator Hamiltonian, $Q_0$ is calculated exactly and in closed form
and it is shown explicitly to be nonunique. The results are verified by using
powerful summation procedures based on analytic continuation. It is also shown
how to calculate the higher coefficients in the perturbation series for $Q$.
\end{abstract}


\maketitle

\section{Introduction}
\label{s1}

The properties of $\cPT$-symmetric Hamiltonians have been observed in a wide
variety of laboratory experiments \cite{R1,R2,R3,R4,R5,R6,R7,R8,R9,R10}. For a
$\cPT$-symmetric Hamiltonian having an unbroken $\cPT$ symmetry a linear
$\cPT$-symmetric operator $\cC$ exists that obeys the following three algebraic
equations:
\begin{eqnarray}
\cC^2&=&1, \label{e1}\\
\left[\cC,\cPT\right]&=&0, \label{e2}\\
\left[\cC,H\right]&=&0. \label{e3}
\end{eqnarray}
Constructing the $\cC$ operator is the key step in showing that time evolution
for a non-Hermitian $\cPT$-symmetric Hamiltonian is unitary \cite{R11,R12}.

The $\cC$ operator for a few nontrivial quantum-mechanical models has been
calculated exactly \cite{R13,R14,R15,R16,R17} by solving (\ref{e1})-(\ref{e3}).
However, in general this system of equations is extremely difficult to solve
analytically. Therefore, in most cases a perturbative approach has been adopted
for the solution of these equations \cite{R18,R19,R20,R21,R22}.

The standard approach to solving (\ref{e1})-(\ref{e3}) has been to express the
$\cC$ operator in a simple and natural form as an exponential of a Dirac
Hermitian operator $Q$ multiplying the parity operator $\cP$:
\begin{equation}
\cC=e^Q\cP.
\label{e4}
\end{equation}
Note that $e^{Q/2}$ is precisely the metric operator $\eta$ discussed in
Refs.~\cite{R23,R24,R25,R26,R27,R28}. This operator can be used to construct a
similarity transformation that maps the non-Hermitian Hamiltonian $H$ to an
isospectral Hermitian Hamiltonian \cite{R29}. If we seek a solution for $\cC$ in
the form (\ref{e4}), we find that (\ref{e1}) and (\ref{e2}), which can be
thought of as {\it kinematical} equations because they hold for all choices of
$H$, imply that $Q(p,q)$ is an odd function of the momentum operator $p$ and an
even function of the coordinate operator $q$ \cite{R11,R12}. The problem is then
reduced to finding the solution to (\ref{e3}), which can be thought of as a {\it
dynamical} equation because it refers to the Hamiltonian $H$.

It is difficult to find a closed-form analytical solution to (\ref{e3}).
However, in the past this equation has been solved perturbatively as follows:
Express the Hamiltonian in the form $H=H_0+\epsilon H_1$ and treat $\epsilon$ as
a small parameter. Then, seek $Q$ as a formal perturbation series in {\it odd}
powers of $\epsilon$:
\begin{equation}
Q(p,q)=\sum_{j=0}^\infty\epsilon^{2j+1}Q_{2j+1}(p,q).
\label{e5}
\end{equation}

The obvious question to ask is, Why do only odd powers in $\epsilon$ appear in
the perturbation series (\ref{e5})? The explanation that has been given in the
past is that even powers of $\epsilon$ are excluded from the series because
$Q(p,q)$ is required to be odd in $p$ and even in $q$. The reasoning goes as
follows: In the quantum-mechanical cases that have been studied so far, such as
$H=H_0+i\epsilon q$ and $H=H_0+i\epsilon q^3$, the unperturbed Hamiltonian $H_0=
\half p^2+\half q^2$ is the harmonic-oscillator Hamiltonian. If there were a
term $Q_0\epsilon^0$ in the series (\ref{e5}), then $Q_0$ would satisfy the
commutation relation $[Q_0,H_0]=0$. The vanishing of this commutator implies
that $Q_0$ is a function of $H_0$, and thus it is an {\it even} function of $p$,
which shows that $Q_0=0$. Once it is established that $Q_0=0$, it is relatively
easy to show (see Ref.~\cite{R15}, for example) that $Q_{2j}=0$ ($j=1,\,2,\,3,\,
\ldots$). We show in this paper that this argument is actually incorrect; there
are indeed solutions to the commutator equation $[Q_0,H_0]=0$ that are {\it odd}
in $p$ -- infinitely many such solutions, in fact. It is precisely because of
the existence of these odd-$p$ solutions that the $\cC$ operator is nonunique.

It has recently become clear that the nonuniqueness of the $\cC$ operator has
important implications for the mathematical and physical interpretation of
$\cPT$-symmetric quantum mechanics \cite{R30,R31,R32}. In Ref.~\cite{R31} it
is shown that if the $\cC$ operator is nonunique, then it is unbounded, and in
this paper we verify this result explicitly.

The paper \cite{R33} is relevant because it discusses for the first time the
existence of multiple (nonunique) solutions to the commutator equation
(\ref{e3}). In Ref.~\cite{R33} it is shown that the {\it inhomogeneous}
commutator equation $[Q_1,H_0]=2H_1$ has an infinite number of particular
solutions, which differ from one another by solutions to the associated {\it
homogeneous} commutator equation $[X,H_0]=0$, and it recognizes that the
solutions $X$ are not necessarily functions of $H_0$ only. The importance of
nonunique solutions to commutator equations is also central to Ref.~\cite{R34},
where a particular time-operator solution $\Theta$ to the inhomogeneous
commutator equation $[\Theta,H]=i$ is called {\it minimal} and a classification
of the infinite number of associated nonminimal solutions is given.

The approach used in the current paper is based on the recognition that finding
multiple solutions for $Q_1$ in (\ref{e5}) is not the only way to demonstrate
nonuniqueness. Here, we introduce a clearer and more fundamental way to explain 
the nonuniqueness of the $\cC$ operator. We show that a more general way to
represent $Q$ is by the expansion
\begin{equation}
Q(p,q)=\sum_{j=0}^\infty\epsilon^jQ_j(p,q)
\label{e6}
\end{equation}
in which all nonnegative integer powers of $\epsilon$ appear. An advantage of
this new representation is that in the limit $\epsilon\to0$ we obtain an
infinite class of {\it exact} $\cC$ operators for the quantum
harmonic-oscillator Hamiltonian $H_0=\half p^2+\half q^2$. Then, once we have
$Q_0$ for the harmonic-oscillator case, we can straightforwardly generalize this
result and verify that the operator $\cC$ is nonunique.

This paper is organized as follows: In Sec.~\ref{s2} we show how to construct
exact and explicit closed-form solutions that are odd in $p$ and even in $q$
to the homogeneous commutator equation $[Q_0(p,q),H_0]=0$. Our result is that
there is a unique bounded $\cC$ operator and a nonunique infinite class of
unbounded $\cC$ operators for the quantum harmonic oscillator. The techniques
used in Sec.~\ref{s2} involve the formal summation of infinite series of
singular operators. However, in Sec.~\ref{s3} we verify the validity of the
formal calculations done in Sec.~\ref{s2} by applying powerful summation
techniques that are used to regulate divergent Feynman integrals. This
verification leads us to conjecture that it may be possible to apply the
principles of summation theory to extend and generalize the rigorous notions of
Cauchy sequences and completeness expansions, which are used in mathematical
Hilbert-space theory, to divergent sequences and series of vectors. Next, in
Sec.~\ref{s4} we develop the formal machinery needed to determine the higher
coefficients $Q_1$, $Q_2$, $\ldots$, in the expansion (\ref{e6}), and in
Sec.~\ref{s5} we concentrate on calculating $Q_1(p,q)$ for the specific case
$H_1=iq$. Finally, in Sec.~\ref{s6} we make some brief concluding remarks.

\section{Solutions to $[Q_0(p,q),H_0]=0$ that are odd in $p$ and even in $q$}
\label{s2}

A powerful strategy for solving operator equations of the form
\begin{equation}
[Q_0(p,q),H_0]=0
\label{e7}
\end{equation}
is to represent the solution $Q_0(p,q)$ as an infinite series of totally
symmetric operator basis functions $T_{m,n}$. The operators $T_{m,n}$ are
described in detail in Refs.~\cite{R34,R35,R36,R37}. However, to make the
presentation in this paper self-contained, we recall that for $m,n\geq0$ the
operator $T_{m,n}$ is defined as a symmetric average over all orderings of $m$
factors of $p$ and $n$ factors of $q$:
\begin{eqnarray}
T_{1,1}&=&\half(pq+qp),\nonumber\\
T_{1,2}&=&\textstyle{\frac{1}{3}}(pqq+qpq+qqp),\nonumber
\end{eqnarray}
and so on. The operators $T_{m,n}$ obey simple commutation and anticommutation
relations:
\begin{eqnarray}
\left[p,T_{m,n}\right]&=&-inT_{m,n-1},\nonumber\\
\left[q,T_{m,n}\right]&=&imT_{m-1,n},\nonumber\\
\left\{p,T_{m,n}\right\}&=&2T_{m+1,n},\nonumber\\
\left\{q,T_{m,n}\right\}&=&2T_{m,n+1},\nonumber\\
\left[p^2,T_{m,n}\right]&=&-2inT_{m+1,n-1},\nonumber\\
\left[q^2,T_{m,n}\right]&=&2imT_{m-1,n+1},
\label{e8}
\end{eqnarray}
where the curly brackets indicate anticommutators. The operator $T_{m,n}$ can be
re-expressed in Weyl-ordered form \cite{R36}:
\begin{equation}
T_{m,n}=\frac{1}{2^m}\sum_{k=0}^m\binom{m}{k}p^kq^np^{m-k}
=\frac{1}{2^n}\sum_{k=0}^n\binom{n}{k}q^k p^m q^{n-k},
\label{e9}
\end{equation}
where $m,\,n=0,\,1,\,2,\,3,\,\cdots$. Introducing the Weyl-ordered form of
$T_{m,n}$ allows one to extend the operators $T_{m,n}$ either to negative values
of $n$ by using the first sum or to negative values of $m$ by using the second
sum. The commutation and anticommutation relations in (\ref{e8}) remain valid
when $m$ is negative or when $n$ is negative.

To find solutions that are odd in $p$ and even in $q$ to the commutator equation
(\ref{e7}), we take $Q_0(p,q)$ to have the general form
\begin{equation}
Q_0^{(\gamma)}(p,q)=\sum_k a_k^{(\gamma)}T_{2\gamma+1-2k,2k},
\label{e10}
\end{equation}
where $\gamma=0,\,\pm1,\,\pm2,\,\ldots$ is a parameter. Substituting (\ref{e10})
into (\ref{e7}), we obtain the following two-term recursion relation for the
coefficients $a_k^{(\gamma)}$:
\begin{equation}
a_{k+1}^{(\gamma)}(k+1)-(\gamma-k+1/2)a_k^{(\gamma)}=0\quad(k=0,1,2,\ldots).
\label{e11}
\end{equation}

This recursion relation is self-terminating; that is, if we choose $a_{-1}^{(
\gamma)}=0$, then $a_0^{(\gamma)}$ is an arbitrary constant, $a_k^{(\gamma)}$
vanishes for $k<0$, and $a_k^{(\gamma)}$ for $k>0$ is determined in terms of
$a_0^{(\gamma)}$ as the solution to the recursion relation (\ref{e11}):
\begin{equation}
a_k^{(\gamma)}=a_0^{(\gamma)}(-1)^k\frac{\Gamma(k-\gamma-1/2)}{k!\Gamma(-\gamma-
1/2)}\quad(k=0,1,2,\ldots).
\label{e12}
\end{equation}
The series (\ref{e10}) with coefficients (\ref{e12}) can be summed as a binomial
expansion:
\begin{equation}
\sum_{k=0}^\infty a_k^{(\gamma)}x^{2k}=a_0^{(\gamma)}\left(1+x^2\right)^{\gamma+1/2}.
\label{e13}
\end{equation}
Thus, for each $\gamma\geq0$ the odd-$p$ and even-$q$ one-parameter family
of solutions to the homogeneous commutator equation (\ref{e7}) is
\begin{equation}
Q_0^{(\gamma)}=\frac{a_0^{(\gamma)}}{2^{2\gamma+2}}\left\{\ldots\left\{\left\{
\left(1+q\frac{1}{p}q\frac{1}{p}\right)^{\gamma+1/2}+\left(1+\frac{1}{p}q\frac{
1}{p}q\right)^{\gamma+1/2},p\right\},p\right\}\dots,p\right\}_{(2\gamma+1)~{\rm
times}},
\label{e14}
\end{equation}
where we have used the identity \cite{R34}
\begin{equation}
T_{-n,n}=\frac{1}{2}\left(q\frac{1}{p}\right)^n+\frac{1}{2}\left(\frac{1}{q}p
\right)^n.
\label{e15}
\end{equation}

As stated in Sec.~\ref{s1}, we can see that while $Q_0^{(\gamma)}$ commutes with
the Hamiltonian $H_0=\half p^2+\half q^2$, it is {\it not} a function of $H_0$
because by construction it is odd in $p$. Furthermore, while the construction of
the solutions in (\ref{e14}) involves series in inverse powers of $p$, these
solutions are well behaved as $p\to0$. To see explicitly the oddness in $p$ we
display the solution corresponding to $\gamma=0$:
\begin{equation}
Q_0^{(0)}=\frac{1}{4}a_0^{(0)}\left(\sqrt{1+q\frac{1}{p}q
\frac{1}{p}}\,\,p+p\,\sqrt{1+q\frac{1}{p}q\frac{1}{p}}+\sqrt{1+\frac{1}{p}q
\frac{1}{p}q}\,\,p+p\,\sqrt{1+\frac{1}{p}q\frac{1}{p}q}\right).
\label{e16}
\end{equation}
In the classical limit for which $p$ and $q$ become commuting numbers, this
solution becomes
\begin{equation}
Q_{0,{\rm classical}}^{(0)}=a_0^{(0)}{\rm sgn}(p)\sqrt{p^2+q^2};
\label{e17}
\end{equation}
the oddness in $p$ is evident.

It is important to point out that for the harmonic oscillator, which corresponds
to $\epsilon=0$, the metric operator $\eta=e^{Q_0}$ is just unity when $Q_0=0$.
Thus, the metric operator is bounded of this special case. However, for $Q_0$ in
(\ref{e16}), the metric operator is no longer bounded, but rather for large $q$
it behaves like $e^q$ and for large $p$ it behaves like $e^p$. Needless to say,
since there is an infinite number of possible choices for $Q_0$, there is an
infinite number of possible metric operators. Only one of the metric operators
is bounded \cite{R38,R39,R40}.

\section{Using summation techniques to verify results of Sec.~\ref{s2}}
\label{s3}

We observed in Sec.~\ref{s2} that even though solutions $Q_0^{(\gamma)}$ in 
(\ref{e14}) were constructed by performing a formal infinite sum over arbitrary
powers of the inverse momentum operator $1/p$, these solutions are well behaved
as $p\to0$. However, the calculations in Sec.~\ref{s2} are certainly not
rigorous. The aim of this section is to provide mathematical support for the
validity of the formulas in (\ref{e14}). Specifically, since the $\cC$ operator
commutes with the Hamiltonian, the $n$th eigenstate $|\psi_n\rangle$ of the 
Hamiltonian must also be an eigenstate of $\cC$. We expect that the eigenvalue
of $|\psi_n\rangle$ is $(-1)^n$. For this to be true, $|\psi_n\rangle$ must be
an eigenstate of $Q_0^{(\gamma)}$ with eigenvalue $0$ for all $n$:
\begin{equation}
Q_0^{(\gamma)}|\psi_n\rangle=0.
\label{e18}
\end{equation}
In this section we show by explicit calculation in which we use powerful
summation techniques that this is indeed the case. Here, we limit our
calculation to the case $\gamma=0$. We first consider the ground state $|\psi_0
\rangle$ and then generalize to the $n$th eigenstate.

From (\ref{e10}) and (\ref{e12}) we see that $Q_0^{(0)}$ is given by
\begin{equation}
Q_0^{(0)}=\sum_{k=0}^\infty a_k T_{1-2k,2k},\qquad a_k=a_0\frac{(-1)^k\Gamma
(k-1/2)}{2\sqrt{\pi}k!}.
\label{e19}
\end{equation}
Also, the unnormalized eigenfunctions of the harmonic-oscillator Hamiltonian in
coordinate space are given by $\psi_n(q)=H_n(q)e^{-q^2/2}$. While the formal sum
in (\ref{e19}) can be written as (\ref{e16}), it is difficult to use this result
to verify the eigenvalue equation (\ref{e18}). A better strategy is to calculate
the action of each term in the sum in (\ref{e19}) on the eigenstates and then to
perform the summation over $k$.

Because inverse powers of the momentum operator arise in (\ref{e19}), it is most
convenient to work in the momentum representation, where the eigenvalue equation
(\ref{e18}) becomes 
\begin{equation}
\langle p|Q_0^{(0)}|\psi_n\rangle=\frac{1}{4}\sum_{k=0}^\infty a_k (-1)^k
\left\{p\left(\frac{1}{p}\partial_p\right)^{2k}+ p\left(\partial_p\frac{1}{p}
\right)^{2k}+\left(\frac{1}{p}\partial_p\right)^{2k}p+\left(\partial_p\frac{1}
{p}\right)^{2k}p \right\}\tilde{\psi}_n(p)=0,
\label{e20}
\end{equation}
where we have used (\ref{e15}) and the third formula in (\ref{e8}). Here,
$\tilde{\psi}_n(p)=(-i)^n\psi_n(p)$, where $\tilde{\psi}$ is the Fourier
transform of $\psi$.

\subsection{Ground state $\psi_0$}
\label{ss3a}

Let we first study the simplest case $n=0$. We can see that the action of each
term in the series expansion (\ref{e20}) on the eigenstate $\psi_0$ produces
more and more negative powers of the momentum $p$: 
\begin{eqnarray}
Q_0^{(0)}(p)\tilde{\psi}_0(p)&=&\frac{1}{4}\left\{4p\,a_0-\left(4p+\frac{2}{p^3}
\right)a_1+\left(4p+\frac{12}{p^3}+\frac{48}{p^5}+\frac{90}{p^7}\right)a_2-
\left(4p+\frac{30}{p^3}+\frac{240}{p^5}+\frac{1350}{p^7}+\frac{5040}{p^9}\right)
a_3\right.\nonumber\\
&+&\left.\left(4p+\frac{56}{p^3}+\frac{672}{p^5}+\frac{6300}{p^7}+\frac{47040}
{p^9}+\frac{264600}{p^{11}}+\frac{997920}{p^{13}}+\frac{1891890}{p^{15}}\right)
a_4-\ldots\right\}e^{-p^2/2}.\nonumber\\
\label{e21}
\end{eqnarray}
We can rearrange the series in (\ref{e21}) to read 
\begin{equation}
Q_0^{(0)}(p)\tilde{\psi}_0(p)=\frac{1}{4}\left\{4p\sum_{k=0}^\infty(-1)^k a_k-
\sum_{k=0}^\infty\frac{(-1)^k}{p^{4k+3}}a_{k+1}P_{2k}(2p^2)\right\}e^{-p^2/2},
\label{e22}
\end{equation}
where $P_n(y)$ are polynomials:
\begin{equation}
P_{n}(y)=\sum_{\alpha=0}^n \frac{(n+1)!(2\alpha+2)!}{2^\alpha \alpha! (\alpha+2)!(n-\alpha)!}y^{n-\alpha}.
\label{e23}
\end{equation}

To verify that $Q_0^{(0)}(p)\tilde{\psi}_0(p)=0$, we must show that
\begin{eqnarray}
4\pi p\sum_{k=0}^\infty\frac{\Gamma(k-1/2)}{k!}-\frac{1}{p^3}\sum_{\alpha=0
}^\infty\frac{1}{p^{2\alpha}}\frac{(2\alpha+2)!}{\alpha!(\alpha+2)!2^\alpha}
\sum_{k=0}^\infty\frac{\Gamma(2k+1)\Gamma(k+3/2)}{\Gamma(k+1)\Gamma(2k-\alpha+1)
}=0,
\label{e24}
\end{eqnarray}
where we have substituted the above formulas for the polynomials $P_n$ and the
coefficients $a_k$. It is easy to verify that the exact sum of the convergent
series $\sum_{k=0}^\infty\Gamma(k-1/2)/k!$ is zero. However, the series
$\sum_{k=0}^\infty\frac{\Gamma(2k+1)\Gamma(k+3/2)}{\Gamma(k+1)\Gamma(2k-\alpha
+1)}$ is divergent for $\alpha\geq3/2$. Therefore, we must introduce a summation
procedure to make sense of this series.

\begin{figure}[b!]
\begin{center}
\includegraphics[scale=0.50]{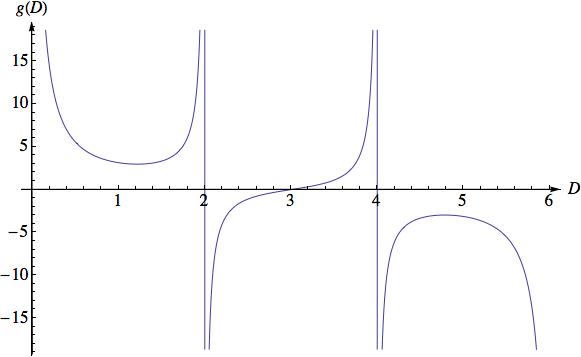}
\end{center}
\caption{A plot of $g(D)$ in (\ref{e26}) for $0<D<6$. Note that $g(3)$ vanishes
even though the integrand of $I(D)$ in (\ref{e25}) is strictly positive for all
$D$. This shows that the notions of positivity and negativity evaporate in the
case of a divergent integral representation.}
\label{F1}
\end{figure}

Our summation procedure is a discrete variant of dimensional continuation, a
technique that is used to interpret divergent Feynman integrals. To illustrate
our approach, let us consider the following $D$-dimensional integral:
\begin{equation}
I(D)=\int d^Dx\frac{x^2+3}{(x^2+1)^2}.
\label{e25}
\end{equation}
This integral converges for $D<2$ and its exact value is $I(D)=S_D\,g(D)$,
where $S_D=2\pi^{D/2}/\Gamma(D/2)$ is the surface area of a $D$-dimensional
sphere of radius $1$ and
\begin{equation}
g(D)=\frac{(3-D)\pi}{2\sin(\pi D/2)}.
\label{e26}
\end{equation}
Evidently, even though the integral representation for $I(D)$ diverges for
$D\geq2$, the function $I(D)$ is well defined and analytic in $D$ except for
isolated simple poles. [Figure \ref{F1} gives a plot of $g(D)$ for $0<D<6$.]
Observe that $I(D)$ vanishes for $D=3$. This result is surprising and somewhat
counterintuitive because the integrand of $I(D)$ is {\it strictly positive} when
$D=3$. The vanishing of $I(3)$ shows that when an integral representation is
divergent we cannot draw qualitative conclusions regarding the sign of its
value. (Indeed, the Borel sum of the series $1+2+4+8+\dots$ is uniquely $-1$
even though all of the terms in this divergent series are positive!)

We will now show that while the sum over $k$ in (\ref{e24}) diverges for $\alpha
>-3/2$, we can evaluate the sum for $\alpha<-3/2$ and then use analytic
continuation in $\alpha$ to sum the series for $\alpha=0,\,1,\,2,\,3,\,\ldots$.
The surprising and counterintuitive result is that while the summand is
positive for $2k+1>\alpha$, the sum of the (divergent) series {\it vanishes} for
all nonnegative integer values of $\alpha$.

\begin{figure}[t!]
\begin{center}
\includegraphics[scale=0.60]{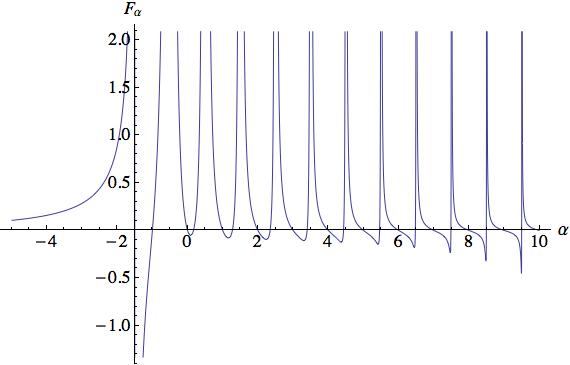}
\end{center}
\caption{A plot of $F_\alpha$ in (\ref{e32}) for $-5<\alpha<10$. Note that
$F_\alpha$ vanishes at every nonnegative integer. This happens because $\Gamma(
-\alpha)$ in the denominator is infinite when $\alpha=0,\,1,\,2,\,\ldots$ and
the hypergeometric function is finite. These zeros are all simple zeros, but the
poles at the half-odd integers beginning with $1/2$ are all double poles, as is
verified in Fig.~\ref{F3}.}
\label{F2}
\end{figure}

The divergent series to be summed is
\begin{equation}
F_\alpha=\sum_{k=0}^\infty\frac{\Gamma(2k+1)\Gamma(k+3/2)}{\Gamma(k+1)\Gamma(
2k-\alpha+1)}.
\label{e27}
\end{equation}
To perform the sum we first express the coefficients in terms of the beta
function: 
\begin{equation}
B(x,y)=\frac{\Gamma(x)\Gamma(y)}{\Gamma(x+y)},
\label{e28}
\end{equation}
whose integral representation is 
\begin{equation}
B(x,y)=\int_0^1 dt\,t^{(x-1)}\,(1-t)^{(y-1)}\qquad({\rm Re}(x),~{\rm Re}(y)>0).
\label{e29}
\end{equation}
By multiplying and dividing $F_\alpha$ by $\Gamma(-\alpha)$ we obtain the result
\begin{equation}
F_\alpha=\frac{1}{\Gamma(-\alpha)}\sum_{k=0}^\infty\frac{\Gamma(k+3/2)}{\Gamma(
k+1)}B(2k+1,-\alpha)=\frac{1}{\Gamma(-\alpha)}\sum_{k=0}^\infty\frac{\Gamma(k+3/
2)}{\Gamma(k+1)}\int_0^t dt\,t^{2k}(1-t)^{-\alpha-1}.
\label{e30}
\end{equation}
We then use the binomial expansion
\begin{equation}
\sum_{k=0}^\infty\frac{\Gamma(k+3/2)}{k!}t^{2k}=\frac{\sqrt{\pi}}{2}(1-t^2)^{-3
/2}
\label{e31}
\end{equation}
to show that 
\begin{equation}
F_\alpha=\frac{\sqrt{\pi}}{2\Gamma(-\alpha)}\int_0^1 dt(1-t)^{-\alpha-5/2}
(1+t)^{-3/2}=\frac{\sqrt{\pi}\,\Gamma(-\alpha-3/2)}{2\Gamma(-\alpha)}{}_2F_1
\left(\frac{3}{2},1;-\alpha-\frac{1}{2},-1\right).
\label{e32}
\end{equation}
This function vanishes for all nonnegative integer values of $\alpha$ because 
the hypergeometric function is finite for these values of $\alpha$. In
Fig.~\ref{F2} we plot $F_\alpha$ for $-5\leq\alpha\leq10$. Observe that
$F_\alpha$ vanishes for $\alpha=0,\,1,\,2,\,\ldots$. Note also that $F_\alpha$
is singular at $\alpha=-3/2$, the value of $\alpha$ for which the series
(\ref{e27}) begins to diverge. Interestingly, this function has {\it double}
poles at the half-odd integers $\alpha=1/2,\,3/2,\,5/2,\,\ldots$. As $\alpha$
increases, the double poles begin to resemble single poles 

\begin{figure}[t!]
\begin{center}
\includegraphics[scale=0.55]{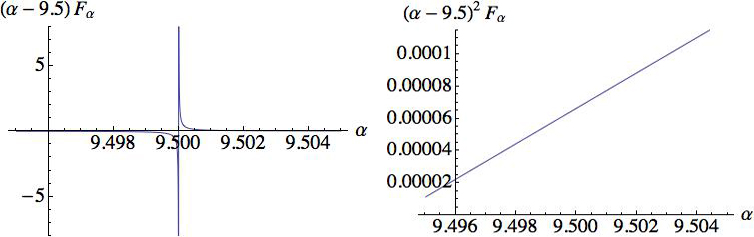}
\end{center}
\caption{A blow-up of the region near $\alpha=9.5$ in Fig.~\ref{F2}. In the left
panel is a plot of $(\alpha-9.5)F_\alpha$ for $9.496<\alpha<9.505$ and in the
right panel is a plot of $(\alpha-9.5)^2F_\alpha$ for the same range of
$\alpha$. Note that the left plot has a typical simple-pole behavior at
$\alpha=9.5$ while the graph in the right panel is finite at $\alpha=9.5$.}
\label{F3}
\end{figure}

\subsection{Generalization: eigenvalue equation for $\psi_n$}
\label{ss3b}

In this subsection we study the action of the operator $Q^{(0)}_0$ on the $n$th
eigenstates of the harmonic-oscillator Hamiltonian. We describe first the case
for which $n$ is even. (The odd-$n$ case is treated in an identical fashion
and is considered briefly at the end of this subsection.) For even $n$
\begin{equation}
Q_0^{(0)}\tilde{\psi}_{2n}=\left[-\sum_{k=1}^\infty \frac{(-1)^k }{p^{4k-1}}a_k
J^{(n)}_{2k-1}(p^2)+\sum_{j=0}^n p^{2j+1}\sum_{k=0}^\infty(-1)^k a_k S^{(n)}_{
n-j}(k)\right]e^{-p^2/2}\quad(n=1,2,3,\ldots).
\label{e33}
\end{equation}
Both the sums over $k$ in (\ref{e33}) diverge for $\alpha=0,1,2,\ldots$ except
for $j=n$ in the first series. (The special case $j=n$ gives the convergent
series that was already considered in Subsec.~\ref{ss3a}.) The divergent series
can be evaluated by using the summation procedure that was introduced in the
previous subsection. Following the procedure for the $n=0$ case, we will show
that if we write (\ref{e33}) as
\begin{equation}
Q_0^{(0)}\tilde{\psi}_{2n}=\left[-\cN^{(n)}+\sum_{j=0}^n p^{2j+1}\cM^{(n,j)}
\right]e^{-p^2/2},
\label{e34}
\end{equation}
then both $\cM^{(n,j)}$ and $\cN^{(n)}$ vanish for $n=1,2,\ldots$, $j=0,1,\ldots
n$. 

We first consider the series
\begin{equation}
\cN^{(n)}= \sum_{k=1}^\infty\frac{(-1)^k }{p^{4k-1}}a_k J^{(n)}_{2k-1}(p^2),
\label{e35}
\end{equation}
where $J^{(n)}_k(x)$ are $k$th order polynomials that can be written as
\begin{equation}
J^{(n)}_k(x)=(2n-1)!!\frac{2^{n+1}}{\sqrt{\pi}}\sum_{\alpha=0}^k\frac{2^\alpha
(k+1)!\Gamma(\alpha+1/2)}{(k-\alpha)!(n+\alpha+1)!}E_{n,\alpha}(k)x^{k-\alpha},
\label{e36}
\end{equation}
and $E_{n,\alpha}(k)$ are polynomials of degree $n$ in the variable $k$. The
first four polynomials are
\begin{eqnarray}
E_{1,\alpha}(k)&=& 2(1-\alpha)k+\alpha^2-4\alpha, \nonumber \\ 
E_{2,\alpha}(k)&=& (4\alpha-8)k^2-(4\alpha^2-20\alpha+4)k+\alpha^3-7\alpha^2+10\alpha, \nonumber \\
E_{3,\alpha}(k)&=& (24-8\alpha)k^3+(12\alpha^2-72\alpha+24)k^2-(6\alpha^2-48\alpha^2+70\alpha-24)k+\alpha^4-9\alpha^3+20\alpha^2-48\alpha,\nonumber\\
E_{4,\alpha}(k)&=& (16\alpha-64)k^4-(32\alpha^2-224\alpha+96)k^3+(24\alpha^3-216\alpha^2+320\alpha-224)k^2\nonumber \\
&&-(8\alpha^4-80\alpha^3+176\alpha^2-536\alpha+96)k+\alpha^5-10\alpha^4+23\alpha^3-158\alpha^2+216\alpha. 
\label{e37}
\end{eqnarray}
In terms of $E_{n,\alpha}(2k-1)=\sum_{\gamma=1}^n e_{\alpha,\gamma}k^\gamma$,
the first term in (\ref{e34}) can be written as
\begin{equation}
\cN^{(n)}=\frac{2}{\pi}(2n-1)!!\sum_{\alpha=0}^\infty \frac{2^\alpha}{p^{2\alpha
+1}}\frac{\Gamma(\alpha+1/2)}{(n+\alpha+1)!}\sum_{\gamma=1}^n e_{\alpha,\gamma}
\,\cN_{\alpha,\gamma},
\label{e38}
\end{equation} 
where $\cN_{\alpha,\gamma}$ is given by
\begin{equation}
\cN_{\alpha,\gamma}=\sum_{k=1}^\infty\frac{\Gamma\left(k-\frac{1}{2}\right)
\Gamma(2k+1)}{\Gamma(k-1)\Gamma(2k-\alpha)}k^\gamma.
\label{e39}
\end{equation}
This is the divergent series that we need to study.

\begin{figure}[t!]
\begin{center}
\includegraphics[scale=0.60]{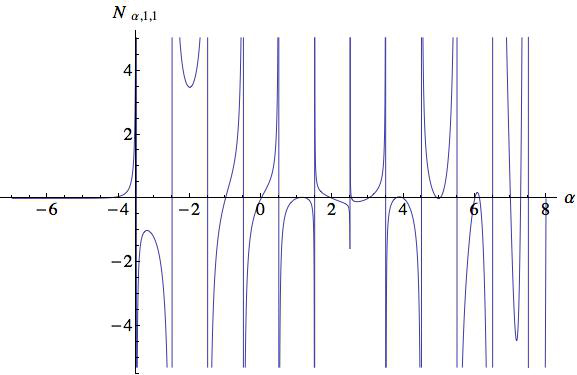}
\end{center}
\caption{A plot of $\cN_{\alpha,1,1}$ in (\ref{e42}) for $-6<\alpha<8$. Like the
special case displayed in Fig.~\ref{F2}, $\cN_{\alpha,1,1}$ vanishes for all
nonnegative integers. However, $\cN_{\alpha,1,1}$ is different from $F_\alpha$
in that it has simple poles rather than double poles. It is not completely
obvious that $\cN_{\alpha,1,1}$ vanishes when $\alpha=0,\,1,\,2,\,\ldots$, so in
Fig.~\ref{F5} the behavior of $\cN_{\alpha,1,1}$ near $\alpha=5$ is blown up.}
\label{F4}
\end{figure}

\begin{figure}[t!]
\begin{center}
\includegraphics[scale=0.60]{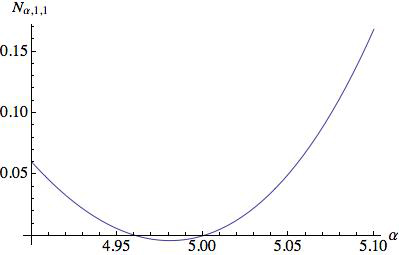}
\end{center}
\caption{A blow up of the graph of $\cN_{\alpha,1,1}$ in Fig.~\ref{F4} near
$\alpha=5$. Observe that there is a zero at exactly $\alpha=5$ as well as an
additional zero near and to the left of $\alpha=5$.}
\label{F5}
\end{figure}

Because $\cN_{\alpha,\gamma}$ is divergent for $\alpha\geq-5/2-\gamma$, we
evaluate the sum for $\alpha<-5/2-\gamma$ and use analytic continuation in
$\alpha$ to sum the series for $\alpha=0,1,2,\ldots$. Multiplying and dividing
$\cN_{\alpha,\gamma}$ by $\Gamma(-\alpha-1)$ and using the integral
representation of the beta function $B$, we obtain
\begin{equation}
\cN_{\alpha,\gamma}=\frac{1}{\Gamma(-\alpha-1)}\int_0^1 dt\,(1-t)^{-\alpha-2}t^2
\sum_{k=0}^\infty \frac{\Gamma(k+1/2)k^\gamma}{\Gamma(k)}t^{2k}.
\label{e40}
\end{equation} 
The sum over $k$ in (\ref{e40}) gives
\begin{equation}
\sum_{k=0}^\infty\frac{\Gamma(k+1/2)k^\gamma}{\Gamma(k)}t^{2k}=\frac{\sqrt{\pi}}
{2}t^4\,{}_{\gamma+1}F_\gamma\left(\frac{3}{2},2,\ldots;1,\ldots;t^2\right),
\label{e41}
\end{equation}
where the first dots in the hypergeometric functions stand for $(\gamma-1)$-twos
and the other dots stand for $(\gamma-1)$-ones. For fixed $\gamma$, the
hypergeometric function in (\ref{e41}) can be written as $L_\gamma(t^2)
2^{-\gamma}(1-t^2)^{-3/2-\gamma}$, where $L_\gamma(t^2)=\sum_{\sigma=1}^\gamma
\ell_{\sigma,\gamma}t^{2\sigma}$ is a polynomial of order $\gamma$ in the
variable $t^2$. In terms of $L_\gamma(t^2)$ the series (\ref{e39}) becomes
$\cN_{\alpha,\gamma}=\sum_{\sigma=1}^\gamma\ell_{\sigma,\gamma}\cN_{\alpha,
\gamma,\sigma}$, where
\begin{eqnarray}
\cN_{\alpha,\gamma,\sigma}&=&\frac{\sqrt{\pi}}{2}\frac{1}{\Gamma(-\alpha-1)}
\int_0^1 dt\,t^{4+2\sigma}(1-t)^{-\alpha-\gamma-7/2}(1+t)^{-\frac{3}{2}-\gamma}
\nonumber\\
&=&\frac{\sqrt{\pi}}{2}\frac{B\left(5+2\sigma,-\alpha-\gamma-\frac{5}{2}\right)}
{\Gamma(-\alpha-1)}{}_2F_1\left(\gamma+\frac{3}{2},5+2\sigma;-\alpha-\gamma+2
\sigma+\frac{5}{2};-1\right).
\label{e42}
\end{eqnarray}
The function $\cN_{\alpha,\gamma,\sigma}$ vanishes for all nonnegative integers
$\alpha$, $\gamma$, and $\sigma$ because the denominator becomes infinite while
the hypergeometric function is finite. The special case $\cN_{\alpha,1,1}$ is
plotted as a function of $\alpha$ in Figs.~\ref{F4} and Fig.~\ref{F5}. The
vanishing of $\cN_{\alpha,\gamma,\sigma}$ guarantees that the first sum in
(\ref{e33}) is identically zero.

Next, we evaluate the sum over $k$ in the divergent series in the second term in
(\ref{e33}):
\begin{equation}
\cM=\sum_{k=0}^\infty (-1)^k a_k S^{(n)}_{n-j}(k),
\label{e43}
\end{equation}
where $S^{(n)}_{n-j}(k)=\sum_{\ell=0}^{n-j}s_{n,\ell}k^\ell$ are polynomials of
degree $n-j$ in the variable k. The polynomials are listed below for $n=1,2,3$:
\begin{eqnarray}
S_1^{(1)}(k)&=& 8(8k+1),\quad S^{(1)}_0(k)=16, \nonumber \\
S^{(2)}_2(k)&=&16(68k^2+14k+3),\quad S^{(2)}_1(k)=-64(8k+3),\quad S^{(2)}_0(k)=
64,\nonumber\\
S^{(3)}_3(k)&=&96(192k^2+52k^2+46k+5),\quad S^{(3)}_2(k)=-64(196k^2+142k+45),
\nonumber\\
&& \,\, S^{(3)}_1(k)=384(8k+5),\quad S^{(3)}_0(k)=-256.
\label{e44}
\end{eqnarray}
As noted above, the series for $j=n$ [the highest power in $p$ in (\ref{e33})]
are convergent and their exact sum is zero. For the divergent cases, the series
(\ref{e43}) can be written as $\cM=\sum_{\ell=0}^{n-j}s_{n,\ell}\cM_\ell$, where
$\cM_\ell$ is the divergent series
\begin{equation}
\cM_\ell=\sum_{k=0}^\infty\frac{\Gamma(k-1/2)}{k!}k^{\ell}.
\label{e45}
\end{equation}
This series can be rewritten in the form
\begin{equation}
\cM_\ell=-\frac{1}{2\sqrt{\pi}}\int_0^1 dt(1-t)^{-3/2} t^{-3/2} \sum_{k=0}^\infty t^k k^\ell.
\label{e46}
\end{equation}

Summing the series in (\ref{e46}) over $k$, we obtain $\sum_{k=0}^\infty t^k
k^\ell=(1-t)^{-\ell-1}\sum_{\sigma=0}^\ell r_{\ell,\sigma}t^\sigma$. Thus,
$\cM_\ell$ can be written as $\cM_\ell=\sum_{\sigma=0}^\ell r_{\sigma,\ell}
\cM_{\ell,\sigma}$, where
\begin{equation}
\cM_{\sigma,\ell}=-\frac{1}{2\sqrt{\pi}}\int_0^1 dt(1-t)^{-5/2-\ell}
t^{-3/2+\sigma}=-\frac{1}{2\sqrt{\pi}}B(\sigma-1/2,-\ell-3/2),
\label{e47}
\end{equation}
which vanishes at zero for $(\sigma,\ell)=0,\,1,\,2,\ldots$.
 
Finally, we consider the case of odd $n$. For this case we get
\begin{equation}
Q_0\tilde{\psi}_{2n+1}=\left[\sum_{j=0}^{n+1}p^{2j}\sum_{k=0}^\infty (-1)^k
a_k V^{(n)}_{n-j}(k)+\sum_{k=0}^\infty \frac{(-1)^k }{p^{4k+2}}a_k O^{(n)}_{2k}
(p^2)\right]e^{-p^2/2}.
\label{e48}
\end{equation}
Here, $V^{(n)}_{n-j}(k)$ are polynomials of degree $n-j$ in the variable k:
\begin{eqnarray}
V_1^{(1)}(k)&=&16k,\quad V^{(1)}_0(k)=-8, \nonumber \\
V^{(2)}_2(k)&=&16(14k^2-k),\quad V^{(2)}_1(k)=-48(4k+1), \quad V^{(2)}_0(k)=32,\nonumber \\
V^{(3)}_3(k)&=&128(26k^3-4k^2+3k),\quad V^{(3)}_2(k)=-32(124k^2+58k+15),
\nonumber\\
&&\,\, V^{(3)}_1(k)=640(2k+1),\quad V^{(3)}_0(k)=-128.
\label{e49}
\end{eqnarray}
Also, the polynomials $O^{(n)}_k(x)$ can be written as
\begin{eqnarray}
O^{(n)}_k(x)=\frac{2^{n+3}}{\sqrt{\pi}}(2n+1)!!\sum_{\alpha=0}^k \frac{2^\alpha\Gamma(\alpha+3/2)(k+2)!}{(k-\alpha)!(n+\alpha+2)!}U_n^{(\alpha)}(k),
\label{e50}
\end{eqnarray}
where $U_n^{(\alpha)}(k)$ are polynomials of degree $n$ in the variable $k$. The
first four such polynomials are
\begin{eqnarray}
U^{(\alpha)}_1(k)&=&1,\nonumber \\ 
U^{(\alpha)}_2(k)&=&2k-\alpha+3,\nonumber \\
U^{(\alpha)}_3(k)&=&4k^2-(4\alpha-12)k+\alpha^2-56\alpha+12,\nonumber\\
U^{(\alpha)}_4(k)&=&8k^3-(12\alpha-36)k^2+(6\alpha^2-30\alpha+76)k-\alpha^3+6\alpha^2-29\alpha+60 . 
\label{e51}
\end{eqnarray}
As in the case of even $n$, we can verify that the sum of both series in
(\ref{e48}) vanish for all $n$.

This completes the verification that the eigenfunction $\psi_n$ of the
harmonic-oscillator Hamiltonian is also an eigenfunction of the $Q_0$ operator
with eigenvalue $0$. Thus, $\psi_n$ is an eigenfunction of $\cC=e^{Q_0}\cP$ with
eigenvalue $(-1)^n$.

\section{Calculation of $Q$ to first order in $\epsilon$}
\label{s4}

The general approach in this paper is to find an operator $\cC=e^Q\cP$ for the
$\cPT$-symmetric Hamiltonian $H=H_0+\epsilon H_1$, where $Q$ has the
power-series expansion (\ref{e6}) in the parameter $\epsilon$ and this expansion
has a nonvanishing zeroth-order term; that is, $Q_0\neq0$. In this section we
concentrate on the formal problem of determining the first-order coefficient
$Q_1$ once $Q_0$ is given.

The coefficient $Q_1$ satisfies the equation
\begin{equation}
\left[e^{Q_0+\epsilon Q_1},H_0\right]=\epsilon\left\{e^{Q_0},H_1\right\},
\label{e52}
\end{equation}
which follows immediately from (\ref{e3}). If we expand this equation to
first order in $\epsilon$, we obtain
\begin{equation}
Z+\frac{1}{2}(Q_0Z+ZQ_0)+\frac{1}{6}(Q_0^2Z+Q_0ZQ_0+ZQ_0^2)+\frac{1}{24}(Q_0^3Z
+Q_0^2ZQ_0+Q_0ZQ_0^2+ZQ_0^3)+\ldots=\left\{e^{Q_0},H_1\right\},
\label{e53}
\end{equation}
where
\begin{equation}
Z\equiv\left[Q_1,H_0\right].
\label{e54}
\end{equation}

Recall that $Q_0$ is a solution to $[Q_0,H_0]=0$, which is a {\it homogeneous}
equation. Thus, any parameter $\mu$ times $Q_0$ is also a solution. Our approach
will now be to make the replacement $Q_0\to\mu Q_0$ in (\ref{e53}) and to treat
$\mu$ as a small perturbation parameter. We can thus expand $Z$ as
\begin{equation}
Z=\sum_{n=0}^\infty Z_n\mu^n
\label{e55}
\end{equation}

To zeroth order in $\mu$ we obtain the result $Z_0=2H_1$. To first order in
$\mu$ we obtain $Z_1=0$, and in fact we find that $Z_{2j+1}=0$ for $j=0,\,1,\,2,
\,\ldots$. The general result for $n\geq2$ can be given in terms of Bernoulli
numbers $\cB_n$:
\begin{equation}
Z_n=\frac{2\cB_n}{n!}\left[Q_0,\ldots\left[Q_0,\left[Q_0,H_1\right]\right]\ldots
\right]_{n~{\rm times}},
\label{e56}
\end{equation}
where $\cB_0=1$, $\cB_1=-\half$ (which is not used in the above formula), and 
$$\cB_2=\frac{1}{6},~\cB_3=0,~\cB_4=-\frac{1}{30},~\cB_5=0,~\cB_6=\frac{1}{42}
,~\cB_7=0,~\cB_8=-\frac{1}{30},~\cB_9=0,~\cB_{10}=\frac{5}{66},~\cB_{11}=0,~
\cB_{12}=-\frac{691}{2730}.$$
We now decompose $Q_1$ into a perturbation series in powers of $\mu$,
\begin{equation}
Q_1=\sum_{n=0}^\infty R_n\mu^n,
\label{e57}
\end{equation}
and obtain a sequence of commutator equations for the coefficients $R_n$:
\begin{equation}
\left[R_n,H_0\right]=Z_n\quad(n=0,\,1,\,2,\,\ldots).
\label{e58}
\end{equation}
In the next section we show how to solve these commutator equations for the
special simple case in which $H_1=q$.

\section{Solution of (\ref{e58}) for the shifted harmonic oscillator
$H=\half p^2+\half q^2+i\epsilon q$}
\label{s5}

Let us consider the shifted harmonic oscillator for which $H_0=\half p^2+\half
q^2$ and $H_1=iq$. This Hamiltonian has an unbroken $\cPT$ symmetry for all real
$\epsilon$. Its eigenvalues $E_n=n+\half+\half\epsilon^2$ ($n=0,\,1,\,2,\,
\ldots$) are all real. One $\cC$ operator for this theory is given exactly by
\cite{R11,R12} 
\begin{equation}
\cC=e^{-2\epsilon p}\cP.
\label{e59}
\end{equation}
In the limit $\epsilon\to0$ the Hamiltonian becomes Hermitian and $\cC$ in
(\ref{e59}) becomes identical with $\cP$. However, the solution for $\cC$ in
(\ref{e59}) is not unique, and by taking any or all of the $Q_0$ in (\ref{e14}),
we obtain an infinite number of operators $\cC$. To find $Q_1$ we must calculate
$Z_0$ (which is $2iq$), $Z_2$, $Z_4$, and so on, and from these we must solve
(\ref{e58}) to obtain $R_0$, $R_2$, $R_4$, and so on.

For the case $n=0$ in (\ref{e58}) we have a simple exact solution to the
commutator equation for $R_0$:
\begin{equation}
R_0=-2p.
\label{e60}
\end{equation}
We emphasize that this solution is not unique.

The equation for $Z_2$ is 
\begin{equation}
Z_2=-\frac{i}{6}\sum_{j=0}^\infty \sum_{k=1}^\infty a_j^{(\gamma)}a_k^{(\gamma)}
(1-2k-2\gamma) F_{j,k}^{(\gamma)},
\label{e61}
\end{equation}
where $F_{j,k}^{(\gamma)}= \left[ T_{-2j-2\gamma+1,2j},T_{-2k-2\gamma,2k}
\right]$, whose explicit form is obtained by using the algebra in
Ref.~\cite{R33} of the basis elements $T_{m,n}$:
\begin{equation}
F_{j,k}^{(\gamma)}=\sum_{\alpha=0}^{j+k-1} \sum_{\beta=0}^{2\alpha+1}
C_{j,k,\alpha,\beta}^{(\gamma)}T_{-2(k+j)-4\gamma - 2 \alpha,2(k+j)-2\alpha-1}
\label{e62}
\end{equation}
with coefficients $C_{j,k,\alpha,\beta}^{(\gamma)}$ given by
\begin{equation}
C_{j,k,\alpha,\beta}^{(\gamma)}=\frac{i^\alpha(-1)^{\alpha+\beta}(2j)!(2k)!(2j+
\beta-2)!(2k+2\alpha-\beta)!}{4^\alpha(2j-2)!(2k-1)!(2j+\beta-2\alpha-1)!(2k-
\beta)!(2\alpha+1-\beta)!\beta!}.
\label{e63}
\end{equation}
A more compact form for $Z_2$ in (\ref{e61}) is
\begin{equation}
Z_2=-\frac{i}{6}\sum_{k=1}^\infty\sum_{\alpha=0}^\infty A_{k,\alpha}^{(\gamma)}
T_{-2k-4\gamma-4\alpha,2k-1},
\label{e64}
\end{equation}
where
\begin{equation}
A_{k,\alpha}^{(\gamma)}=2\frac{(-1)^{k-\alpha+1}\Gamma(k+2\alpha+2\gamma)
\Gamma^2(\alpha+\gamma+1/2)}{\Gamma(\gamma-1/2)\Gamma(k)\Gamma(\alpha+1)\Gamma
(\alpha+2\gamma)(\alpha+\gamma)}
\label{e65}
\end{equation}
and for simplicity we have set $a_0=1$.

For the special case $\gamma=0$ we have the following results for $R_2^{(0)}$,
where from now on we omit the superscript $(0)$. The commutator equation is
\begin{equation}
\left[R_2,\,H_0\right]=\frac{i}{6}\sum_{k=1}^\infty\sum_{\alpha=0}^\infty
A_{k,\alpha}T_{-2k-4\alpha,2k-1}.
\label{e66}
\end{equation}
This is a linear equation, so we solve it for each $\alpha$ separately and
express the solution as a sum over $\alpha$: $R_2=\sum_{\alpha=0}^\infty\rho_{
k,\alpha}R_{2,\alpha}$. For $\alpha=0$ we seek a solution of the form
\begin{equation}
R_{2,0}=\sum_{k=0}^\infty\rho_{k,0}T_{-2k-1,2k}
\label{e67}
\end{equation}
whose coefficients $\rho_{k,0}$ satisfy the recursion relation
\begin{equation}
(2k-1)\rho_{k-1,0}+2k\rho_{k,0}=A_{k,0}.
\label{e68}
\end{equation}
(The techniques used here are described in detail in Ref.~\cite{R33}.) The
simplest solution to this recursion relation is $\rho_{k,0}=\pi(-1)^k$.

For $\alpha=1$ we set
\begin{equation}
R_{2,1}=\sum_{k=0}^\infty\rho_{k,1}T_{-2k-5,2k}
\label{e69}
\end{equation}
and so the recursion relation for the coefficients $\rho_{k,1}$ is
\begin{equation}
(2k+3)\rho_{k-1,1}+2k\rho_{k,0}=A_{k,1},
\label{e70}
\end{equation}
whose solution is $\rho_{k,0}=-\frac{\pi}{4}(-1)^k(k+2)!/k!$.

For general $\alpha$ we have
\begin{equation}
R_{2,\alpha}=\sum_{k=0}^\infty\rho_{k,\alpha}T_{-2k-4\alpha-1,2k}
\label{e71}
\end{equation}
and the recursion relation for the coefficients $\rho_{k,\alpha}$ is
\begin{equation}
(2k+4\alpha-1)\rho_{k-1,\alpha}+2k\rho_{k,\alpha}=A_{k,\alpha},
\label{e72}
\end{equation}
whose solution is
\begin{equation}
\rho_{k,\alpha}=(-1)^k\frac{(k+2\alpha)!}{k!}\left[\frac{\Gamma(\alpha+1/2)}
{\alpha!}\right]^2.
\label{e73}
\end{equation}

\subsection{Complete evaluation of the first-order expansion $Q_1$}
\label{ss5a}

We now derive the general form of the first-order expansion in $\epsilon$ of the
operator $Q=Q_0+\epsilon Q_1+\epsilon^2 Q_2+\ldots$, which takes the form of a
series in even powers of the parameter $\mu$:
\begin{equation}
Q_1=\sum_{n=0}^\infty\mu^{2n}R_{2n},
\label{e74}
\end{equation}
whose coefficients $R_{2n}$ satisfy the commutator equation
\begin{equation}
\left[R_{2n},H_0\right]=Z_{2n}\quad(n=1,2,3,\ldots).
\label{e75}
\end{equation}
Recall that the operator $Z_{2n}$ is proportional to the recursive evaluation of
the double commutator $[Q_0,[Q_0,\,.\,]]$ acting on the operator $Z_{2n-2}$,
starting from $Z_0=2H_1$. (As established earlier, $Z_2=[Q_0,[Q_0,H_1]]/6$, $Z_4
=-[Q_0,[Q_0,[Q_0,[Q_0,H_1]]]]/360$, and so on.) The operator $Z_{2n}$ can be
written as a double series over the basis elements $T_{m,n}$ once their algebra
has been repeatedly applied and its closed-form expression is
\begin{equation}
Z_{2n}=2i\frac{\cB_n}{n!}\sum_{k=n}^\infty\sum_{\alpha=0}^\infty A^{(2n)}_{k,
\alpha}T_{-2k-4\alpha,2k-2n+1}\quad(n=1,2,3,\ldots),
\label{e76}
\end{equation}
where the coefficients $A^{(2n)}_{k,\alpha}$ are given by
\begin{equation}
A^{(2n)}_{k,\alpha}=W_\alpha^{(2n)}(-1)^k\Gamma(k+2\alpha+n-1)/\Gamma(k).
\label{e77}
\end{equation}

It is extremely laborious to evaluate explicitly the function $W_\alpha^{(2n)}$,
even for the first few values of $n$. For example, for $n=1$ we have $W_\alpha^{
(1)}=[\Gamma(\alpha+1/2)/\alpha!]^2$; the evaluation of the next commutator with
$Q_0$ gives
$$W_\alpha^{(2)}=\sum_{\beta=0}^\alpha\frac{\Gamma(2\alpha-\beta+1)\Gamma(\beta+
1/2)\Gamma^2(\alpha-\beta+1/2)}{(4\alpha+1)\beta!\Gamma(2\alpha-\beta+1/2)
\Gamma^2(\alpha-\beta+1)}.$$
Moreover, the existence of solutions $R_{2n}$ to (\ref{e75}) is not affected by
the explicit form of the functions $W_\alpha^{(2n)}$ because (\ref{e75}) is a
linear equation. 

Substituting (\ref{e76}) into (\ref{e75}) and noting the last two commutator
equations in (\ref{e8}), we argue that the operator $R_{2n}$ has the form
\begin{equation}
R_{2n}=\sum_{k=n-1}^\infty\,\sum_{\alpha=0}^\infty\rho^{(2n)}_{k,\alpha}
T_{-2k-4\alpha-1,2k-2n+2},
\label{e78}
\end{equation}
where the coefficients $\rho_{k,\alpha}^{(2n)}$ satisfy the recursion relation
\begin{equation}
(2k+4\alpha-1)\rho_{k-1,\alpha}^{(2n)}+2(k-n+1)\rho_{k,\alpha}^{(2n)}=A_{k,
\alpha}^{(2n)}.
\label{e79}
\end{equation}
Equation (\ref{e79}) is a first-order linear inhomogeneous difference equation
that we can rewrite as
\begin{equation}
\rho_{k+1,\alpha}^{(2n)}+\frac{2k+4\alpha+1}{2(k-n+2)}\rho_{k,\alpha}^{(n)}=\frac{A_{k+1,\alpha}^{(2n)}}{2(k-n+2)}.
\label{e80}
\end{equation} 

To find solutions to (\ref{e80}) we divide both sides of the equation by the
{\it summing factor} $Y_k$, where 
\begin{equation}
Y_k=(-1)^k\prod_{j=n}^k\frac{(2j+4\alpha+1)}{2(j-n+2)}=\frac{(-1)^k\Gamma(k+2
\alpha+3/2)}{\Gamma(2\alpha+n+1/2)\Gamma(k-n+3)}.
\label{e81}
\end{equation}
Equation (\ref{e80}) then becomes
\begin{equation}
\frac{\rho_{k+1,\alpha}^{(2n)}}{Y_k}-\frac{\rho_{k,\alpha}^{(2n)}}{Y_{k-1}}=
\frac{A_{k+1,\alpha}^{(2n)}}{2(k-n+2)Y_k}.
\label{e82}
\end{equation} 
Note that (\ref{e82}) has taken the form of an exact discrete difference of the
function $\rho_{k,\alpha}^{(2n)}/P_{k-1}$. Summing both sides of (\ref{e82})
from $1$ to $k-1$ gives the solution
\begin{equation}
\rho_{k,\alpha}^{(2n)}=\frac{(-1)^{k+1}\Gamma(k+2\alpha+1/2)}{\Gamma(n+2\alpha+
1/2)(k-n+1)!}\left[G_\alpha^{(2n)}+W_\alpha^{(2n)}\sum_{j=1}^{k-1}\frac{\Gamma
(n+2\alpha+1/2)(j+2\alpha+n-2)!(j-n+1)!}{2\,\Gamma(j+2\alpha+3/2)(j-1)!}\right],
\label{e83}
\end{equation}
where $G_\alpha^{(2n)}$ is an arbitrary constant.

\subsection{Semiclassical approximation to the $\cC$ operator}
\label{ss5b}

In this subsection we attempt a semiclassical calculation of the $\cC$ operator.
In such a calculation we expand $Q$ as a series in powers of $\hbar$:
\begin{equation}
Q=Q_0+\hbar\,Q_1+\hbar^2 Q_2+\ldots.
\label{e84}
\end{equation}
The semiclassical approximation terminates after the $\hbar$ term in this
expansion. The ordering of powers of $p$ and $q$ in this expansion becomes
unimportant because commuting $p$ with $q$ introduces additional powers of
$\hbar$. Furthermore, every factor of $pq$ has dimensions of $\hbar$, and thus
only the first term in a sum needs to be kept.

A recursive determination of the first-order solution $Q_1$ in (\ref{e84})
arises as a natural simplification of the difficult problem that we formally
solved in Subsec.~(\ref{ss5a}). To proceed, a dimensional analysis of the
operators is required. Because of the commutator equation $[q,p]=i\hbar$, we can
assign the dimensions of $q$ and $p$ to be $\hbar^{1/2}$. With this convention
the Hamiltonian for the shifted harmonic oscillator becomes
\begin{equation}
H=\half p^2+\half q^2+i\hbar^{1/2}q.
\end{equation}
\label{e85}

For $Q_0$ in (\ref{e19}) to be the zeroth-order solution in $\hbar$, we must
introduce its explicit dependence on $\hbar$; that is 
\begin{equation}
Q_0=\hbar^{-1/2}\sum_{k=0}^\infty a_k T_{1-2k,2k}.
\end{equation}
\label{e86}
Following the procedure illustrated in Sec.~(\ref{s4}), we make the replacement
$Q_0\to\mu\,Q_0$, where $\mu$ is a small {\it dimensionless} parameter. The
operator $Q_1$ in (\ref{e74}) admits dimensionless solutions in terms of the
operators $R_{2n}$ only for $\alpha=0$ in (\ref{e78}). In fact, for $\alpha=0$
the series representation of both $Z_{2n}$ in (\ref{e76}) and $R_{2n}$ in
(\ref{e78}) can be drastically simplified. Introducing the explicit dependence
on $\hbar$, the operator $Z_{2n}$ becomes
\begin{equation}
Z_{2n}=\hbar^{n+1/2}\sum_{k=n}^\infty\frac{(-1)^{n+k-1}(k-1)!}{(n-1)!(k-n)!}
T_{-2k,2k-2n+1},
\label{e87}
\end{equation}
while for the operators $R_{2n}$ we get
\begin{equation}
R_{2n}=\hbar^{n-1/2}\sum_{k=n-1}^\infty\rho_k^{(2n)}T_{-2k-1, 2k-2n+2},
\label{e88}
\end{equation}
where the coefficients $\rho_k^{(2n)}$ satisfy the recursion relation
\begin{equation}
\rho_{k+1}^{(2n)}+\frac{2k+1}{2(k-n+2)}\rho_k^{(2n)}=\frac{(-1)^{n+k}k!}{2(n-1)!(k-n+2)!}.
\label{e89}
\end{equation}

The general solution to (\ref{e89}) contains an arbitrary constant $C_n$:
\begin{equation}
\rho_k^{(2n)}=C_n+\frac{(-1)^n\Gamma(n+1/2)[\sqrt{\pi}\,k!-\Gamma(k-1/2)]}{
\sqrt{\pi}(n-1)!\Gamma(k+1/2)}.
\label{e90}
\end{equation}
With the choice $C_{n}=(-1)^n\pi^{-1/2}\Gamma(n+1/2)/(n-1)!$ the result in
(\ref{e90}) is considerably simplified. The simplest first-order solution $Q_1$
in the series (\ref{e84}) for the operator $Q$ is
\begin{equation}
Q_1=\hbar^{-1/2}\sum_{n=0}^\infty\frac{(-\mu^2\hbar)^n}{(n-1)!}
\sum_{k=n-1}^\infty\frac{k!}{\Gamma(k+1/2)}\,T_{-2k-1,2k-2n+2}.
\end{equation}
\label{e91}
This illustrates the nature of a semiclassical expansion for the operator $Q$.

\section{Conclusions}
\label{s6}

The principal result in this paper is that while there is a unique bounded
metric and $\cC$ operator for the quantum harmonic oscillator, which is the
simplest $\cPT$-symmetric quantum theory, there is an infinite number of
unbounded metric and $\cC$ operators, and we have calculated them exactly.
To produce these unbounded operators we have had to sum infinite series of
singular operators (involving powers of $1/p$) and have observed that the
resulting sums are no longer singular. Of course, our summation procedure is at
best only formal. However, we have verified our results by using dimensional
summation procedures and have shown that the $\cC$ operators that we have
constructed satisfy exactly their defining equations.

As anticipated in Ref.~\cite{R31}, the properties of nonuniqueness and
unboundedness of the $\cC$ operators are connected. There is a unique bounded 
$\cC$ operator for the harmonic oscillator, namely $\cP$, and an infinite class
of unbounded $\cC$ operators. Interestingly, the unbounded metrics grow for
large $q$ like $e^q$. This does not pose a serious problem if we want to
calculate matrix elements of eigenstates of $H_0$ because in $q$ space these
states vanish like $e^{-q^2}$. Thus, any finite linear combination of
eigenstates is an acceptable state in the Hilbert space associated with the
unbounded metric.

Finally, while we have performed formal summations of operators in this paper,
we have justified our results by doing careful summation calculations that rely
on analytic continuation. Our calculation are modeled on the dimensional
continuation evaluations that are used to regulate divergent Feynman integrals.
We conjecture that such techniques might be applied to generalize the notions of
Cauchy sequences and completeness sums for Hilbert-space vectors.

\acknowledgments
We thank S.~Kuzhel for many discussions regarding Hilbert-space theory and
Q.~Wang for useful comments regarding Sec.~\ref{s3}. MG is grateful for the
hospitality of the Department of Physics at Washington University. CMB thanks
the U.S.~Department of Energy and the U.K.~Leverhulme Foundation and MG thanks
the INFN (Lecce) for financial support.

\end{document}